# Rhodium Nanoparticles for Ultraviolet Plasmonics

*Anne M. Watson[1], Xiao Zhang[2], Rodrigo Alcaraz de la Osa[3], Juan Marcos Sanz[3], Francisco G. Fernández[3], Fernando Moreno[3], Gleb Finkelstein[1], Jie Liu[2], and Henry O. Everitt[1,4] **

1 Department of Physics, Duke University, Durham, North Carolina 27708, United States

2 Department of Chemistry, Duke University, Durham, North Carolina 27708, United States

3 Department of Applied Physics, University of Cantabria, Santander, Cantabria 39005, Spain

4 Army, Aviation & Missile RD&E Center, Redstone Arsenal, Alabama 35898, United States

Abstract

The non-oxidizing catalytic noble metal rhodium is introduced for ultraviolet plasmonics. Planar tripods of 8 nm Rh nanoparticles, synthesized by a modified polyol reduction method, have a calculated local surface plasmon resonance near 330 nm. By attaching p-aminothiophenol, local field-enhanced Raman spectra and accelerated photo-damage were observed under near-resonant ultraviolet illumination, while charge transfer simultaneously increased fluorescence for up to 13 minutes. The combined local field enhancement and charge transfer demonstrate essential steps toward plasmonically-enhanced ultraviolet photocatalysis.

In the growing field of nanoplasmonics, metal nanoparticles (NPs) are routinely used to alter the electromagnetic field from an external irradiation source to produce strong local field effects through the local surface plasmon resonance (LSPR).[1–4] Molecular analytes in the vicinity of these metal NPs experience enhanced spectral, photocatalytic, and/or photodegradation responses, especially when excited near the LSPR energy.[5–12] Nanoplasmonics research has traditionally used copper and the noble metals gold and silver, which constrain their operation to the visible or near infrared (NIR) spectral regions. Although Ag has superior plasmonic performance, the oxide that eventually forms upon exposure to air tarnishes its effectiveness and limits the enhancement felt by a nearby analyte. Copper also tarnishes upon exposure to air, so Au, which forms no native oxide and rarely tarnishes, has become the most widely used nanoplasmonic metal in these spectral regions.

Applications for nanoplasmonics are extending into the ultraviolet (UV) spectral region.[13–21] Because Ag, Au, and Cu cannot operate in the UV, the search is on for new metals for UV nanoplasmonics. Most of the recent work has focused on aluminum or gallium, both of which are compelling because of their low cost, wide availability, high conductivity, lack of UV interband transitions, and compatibility with CMOS processing.[13,15,17–19,22,23] However, Al suffers from the formation of a native oxide layer several nanometers thick, as do other compelling UV plasmonic metals including indium, magnesium, titanium, tin, thallium, and lead.[16,20] Gallium and chromium are more attractive UV plasmonic metals because their native oxide is only a few monolayers thick, but oxide-free noble metals with UV plasmonic performance would be even more desirable. Of these noble metals - Pt, Pd, Ru, and Rh - a recent theoretical study has found that only Rh has a strong UV plasmonic response, and it is possible to fabricate Rh nanoparticles smaller than 10 nm through chemical means.[20,24] Moreover, Rh is routinely used as a hydrogenation catalyst for a wide variety of alkenes, alkynes, and aromatic cyclic arenes, as well as nitriles, pyridines, and various N-heterocyles. Indeed, the primary industrial use of Rh is in three-way catalytic converters to reduce $NO_x$, where Rh is often alloyed with Pt and Pd because of its corrosion resistance.[25] Given the rapidly growing interest in plasmonically-assisted photocatalytic processes, Rh represents a tremendously promising metal for combining plasmonics and catalysis.

Here we report the UV plasmonic properties of Rh NPs by means of surface-enhanced Raman spectroscopy, surface-enhanced and charge transfer-enhanced fluorescence, and photo-induced degradation. Rh tripod structures with 8 nm-long arms were chemically synthesized, and surface-enhanced spectroscopy was performed on p-aminothiophenol (PATP) attached to the surface of the Rh nanostructures. Comparing the response for laser excitation at UV and visible wavelengths (resonant and not resonant with the Rh NP LSPR, respectively), we found that Raman and fluorescence spectra were enhanced and photo-degradation was accelerated in the presence of Rh under resonant UV excitation. Raman spectra usually decayed rapidly upon UV exposure, an indication of locally enhanced photodamage of PATP. However, in some locations the Raman spectra were stable and the fluorescence spectra increased, often for many minutes, an indication that photoexcited hot electrons efficiently transferred from the Rh nanostructure to the PATP before photodamage ultimately quenched both spectral features. These findings confirm the exciting potential of Rh nanostructures for UV plasmonic and photocatalytic applications.

We used Rh NPs of a tripod geometry, synthesized following a modified polyol reduction method (see Methods section for complete synthesis details).[24] $RhCl_3 \cdot xH_2O$ and PVP were combined and heated to produce the desired reaction. The product was collected by centrifugation and washed by water and acetone several times. The resultant PVP-coated black Rh tripods were then dispersed in ethanol for characterization. TEM images of the Rh NPs such as in Figure 1a indicated a relatively monodisperse distribution of tripods with representative arm lengths of ~8 nm and widths of ~3.5 nm, while AFM imaging established that their maximum height was 8-10 nm.

Figure 1b compares the absorption spectra of Rh NPs randomly dispersed in ethanol as calculated by the discrete dipole approximation (see Methods section)[26] and as measured by a UV-vis absorption spectrometer. Considering only the representative NP size and shape was modeled for random orientations, the theoretically predicted peak for the dipolar mode in the tripod plane near 3.3 eV (375 nm) is in good agreement with the experimental data and indicates that ethanol redshifted the Rh LSPR energy by more than 0.4 eV from 3.74 (332 nm) for Rh NPs in air. A stronger peak near 4.6 eV is also predicted for the dipolar mode oriented perpendicular to the tripod plane, and it was excited for randomly oriented NPs in ethanol but with a narrower linewidth that reflects a sensitivity of this mode to NP height. Although this perpendicular mode cannot be excited for bare NPs deposited on a planar substrate using normal incidence illumination, calculations predict that the in-plane mode blueshifts to 3.47 eV (357 nm) with a linewidth of 0.5 eV FWHM. Consequently, for the surface-enhanced spectroscopy measurements in this configuration, the LSPR nearly coincides with the UV laser excitation at 3.82 eV (325 nm) and is well removed from the visible excitation at 2.54 eV (488 nm). COMSOL calculations of the near field $|E|^2$ indicate the field enhancements for Rh NPs on a 1 μm-thick $SiO_2$ "glass" substrate are strongest near the tips of the tripods, as illustrated for two representative polarization directions in Figure 1c.[27] The calculated LSPR peak was found to be relatively insensitive to whether the Rh NP arms were wedge-shaped or conical, but the LSPR strengthened and blueshifted significantly with increasing height.[28]

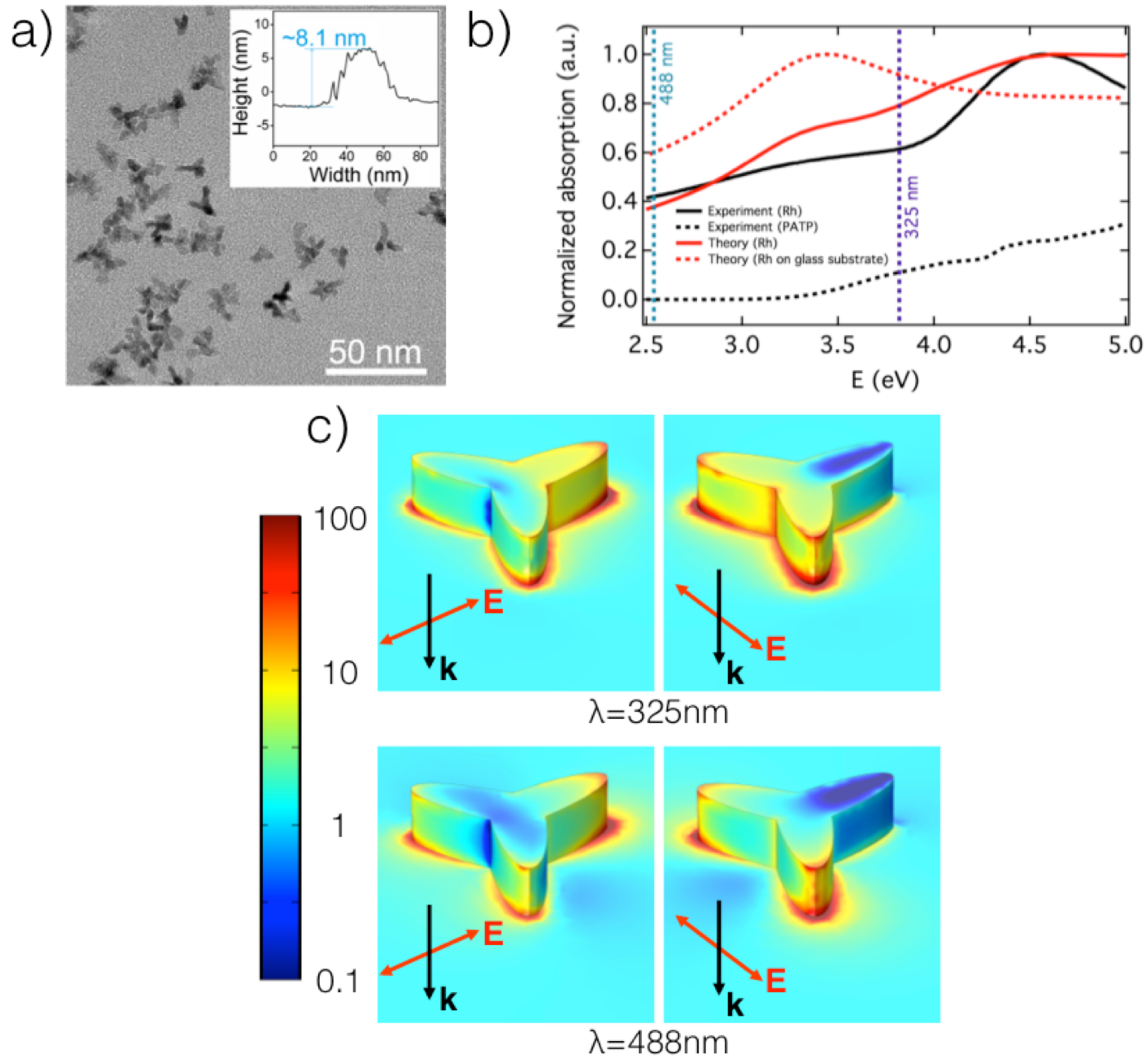

**Figure 1**: Characterization of the Rh NPs. (a) TEM image showing Rh NP tripods, with inset showing the AFM-measured height profile. (b) UV-vis absorption spectra experimentally obtained from the Rh NPs in ethanol solution (black solid) and PATP (black dotted), overlaid by the calculated spectra for Rh NPs in ethanol (red solid) and on an $SiO_2$ "glass" substrate (red dotted). The UV and visible laser lines are also shown. (c) Near field intensity $|E|^2$ for Rh NPs on a glass substrate illuminated at normal incidence. Both laser wavelengths and two representative linear polarization directions are shown.

To characterize the plasmonic response of the Rh NPs under UV and visible excitation, samples were prepared on 2x2 mm silicon substrates, covered by a 1 μm-thick $SiO_2$ layer, to facilitate SEM imaging. They were cleaned using the standard RCA method followed by a plasma ashing process to ensure a clean sample. A high concentration of the Raman dye molecule PATP, prepared in a 5 mM ethanol solution, was used to ensure a strong spectroscopic signal. For the reference Raman measurements (PATP only), a 10 μL sample of this solution was directly drop-cast onto the silicon substrate. For the Rh+PATP samples, the Rh NPs were etched in nitric acid to remove the PVP capping agent, then washed with ethanol. Because thiol linkers strongly attach to bare metallic surfaces, 150 μL of the 5 mM PATP solution was added to 50 μL of this washed Rh NP solution, causing a monolayer of PATP to attach to the surfaces of the Rh NPs. The resultant Rh+PATP solution was allowed to incubate in a covered shaker at room temperature for 3 hours, then was left undisturbed for 12 hours before being centrifuged 10 min at 2g using a filter to remove the excess un-attached PATP. Finally, the Rh+PATP residual was drop-cast onto the substrates such that only the surfaces of the Rh NPs were coated with PATP. All samples were blown gently with $N_2$ after a 10 min incubation period to complete the drying process.

For both UV and visible excitation sources, the samples consisting only of Rh NPs (no PATP) exhibited Raman spectra only from the substrate, indicating that any remaining PVP did not contribute to the spectra. Any additional signal from the Rh+PATP samples comes from the attached PATP, so the spectra taken from the PATP reference sample and spectra from the Rh+PATP sample may be directly compared to reveal the enhancement caused by the Rh NPs. The Rh+PATP NPs were coated by a 0.7 nm-thick monolayer of PATP,[29]and SEM imagery indicated the NPs covered ~3% of the illuminated area. Given that the reference sample was coated by ~10 nm of PATP (as determined by AFM imaging), the volumetric ratio of PATP in the reference versus the Rh+PATP NPs is estimated to be about 200:1, a geometric mean for the range of conical (500:1) and wedge-shaped (50:1) arms of the representative NP.

Comparing the black curves in Figure 2, it can be seen that different Raman modes from the $Si/SiO_2$ substrate (located at either 520 or 860 $cm^{-1}$)[30,31] were excited under UV and visible excitation, and both were stable over time. The measured PATP Raman spectra were scaled to the height of the strongest observed substrate mode so that the enhancement factors for different excitation wavelengths may be compared quantitatively. The vibrational modes excited in the PATP molecule included the C-S stretch (1080 $cm^{-1}$) and C-C stretch (1600 $cm^{-1}$) modes, which are in-plane modes of the $a_1$ type.[32] Both modes appeared in the spectra for both excitation wavelengths, while a broad band of modes between 1200-1300 $cm^{-1}$ appeared for UV excitation but was obscured by a substrate feature for visible excitation. Our analysis therefore concentrates on the C-C stretch mode because it was clearly distinguishable in all samples. The C-S stretch mode, which was also frequently but weakly visible, decayed within ~60 sec, so it was not analyzed here.

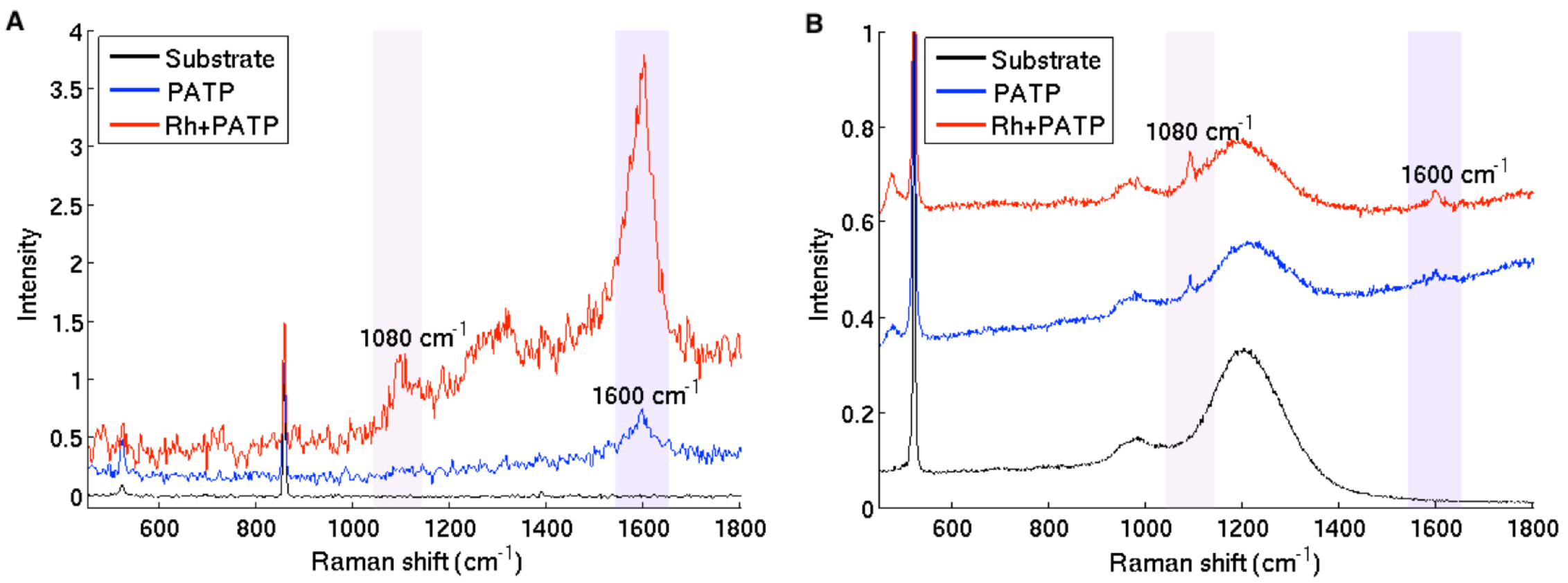

**Figure 2**: Representative spectra for the silicon substrate, PATP alone, and Rh coated with PATP (Rh+PATP) under (a) 325 and (b) 488 nm excitation. The spectra are normalized using the dominant substrate peak and offset vertically for clarity. The PATP peaks at 1080 $cm^{-1}$ and 1600 $cm^{-1}$ are highlighted.

Figure 2 compares the Raman spectra without (blue) and with (red) Rh NPs, excited by the UV and visible lasers. The reference Raman spectra for pure PATP are stronger under UV excitation than for visible excitation because the scattering cross section scales as the fourth power of frequency (factor of ~5x) and because excitation at 325 nm weakly overlaps the tail of the PATP absorption band (Fig. 1b).[32] The striking enhancement of the $a_1$ C-S and C-C stretch modes at 1080 and 1600 $cm^{-1}$ under resonant UV excitation of the Rh+PATP sample is a clear manifestation of local field enhancement produced by the UV plasmonic properties of Rh NPs. If charge transfer alone was responsible for enhancing the spectra, the $b_2$ modes at 1142, 1391, and 1440 $cm^{-1}$ should have been more strongly enhanced than the $a_1$ modes.[32–34] However, charge transfer enhancement depends strongly on the orientation of the PATP on the surface. Since the molecule is oriented nearly perpendicular to the surfaces and the largest surface area of the NPs is the sidewalls, the vast majority of the PATP is oriented horizontally in a manner that is unfavorable for the $b_2$ modes to be seen.[32,34,35]

Quantitatively examining the C-C mode strength, the enhancement produced by the Rh NPs for UV excitation is about a factor of 4.5. Assuming a PATP volume ratio of 200, the estimated local field enhancement of the Raman spectra is $\sim 10^3$, and Figure 1c indicates that the enhancement is much greater at the tip(s) most closely aligned with the polarization of the laser. By contrast, under visible excitation the Raman spectra were enhanced by only a factor of $<1.5$, which is weaker by a factor of ~3 than the measured UV enhancement. This UV/visible enhancement ratio is in excellent agreement with the theoretically predicted enhancement ratio (3.2) when averaged over all orientations of the Rh tripods relative to a fixed linear polarization of the lasers.

Figure 3 plots the Raman spectra for a PATP + Rh NPs sample, accumulated by repeated 20 and 30 s exposures for either UV or visible illumination of a single spot, respectively. The $a_1$ C-C stretch mode at 1600 $cm^{-1}$ decays over time with a rate that depends on whether or not the Rh NPs are present and whether or not the excitation is resonant with the LSPR. Like most organic molecules, PATP experiences greater photo-degradation upon exposure to UV illumination than to visible illumination. Moreover, when the Rh NPs are excited at 325 nm near the LSPR resonance, regions of intense electric field are created near the surface of the Rh particles. PATP molecules in this intense field experience accelerated photo-damage, and their contribution to the Raman spectrum is lost.[17,36,37]

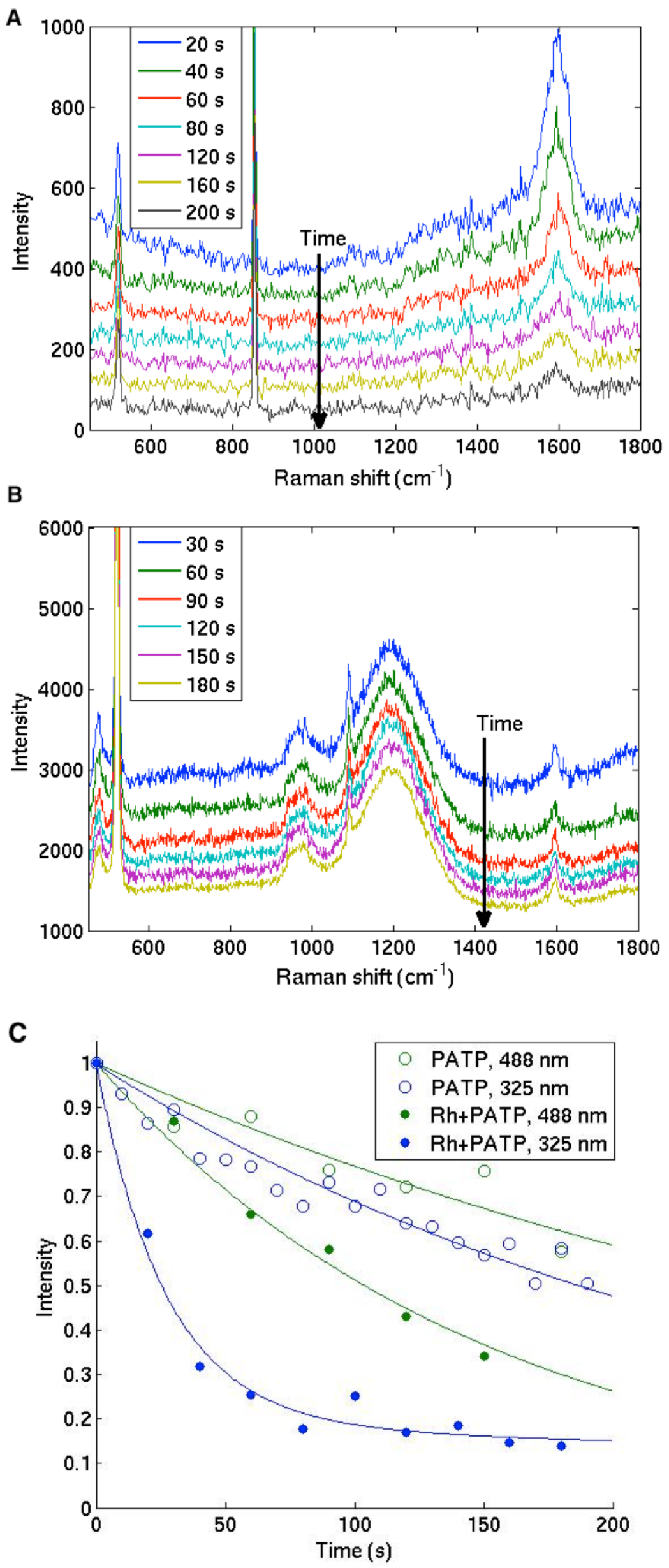

**Figure 3**: Representative, temporally evolving Raman spectra for the Rh+PATP samples for a single illumination spot under (a) UV (325 nm) and (b) visible (488 nm) excitation. Spectra are offset vertically for clarity, and collection times are cumulative. (c) Temporal dependence of the 1600 cm$^{-1}$ peak for the Rh+PATP samples, compared to the PATP reference sample, for both excitation wavelengths. Data is scaled such that the fitted curves are normalized to 1 at time t = 0.

Thus, the rate of Raman signal decay depends both on the excitation wavelength and its coincidence with the NP LSPR. We can evaluate the relative contribution of these two mechanisms by examining the decay rates for the Rh+PATP samples and the PATP reference samples for both excitation wavelengths. As compared to the PATP reference sample, the faster decay rate for the Rh+PATP (Figure 3c) for resonant excitation is a sign of plasmon-accelerated photo-degradation. Although the same sample under non-resonant 488 nm illumination also decays at an accelerated pace compared to the pure PATP sample, the accelerated decay is less pronounced than for the intense local fields generated with resonant LSPR excitation. Fitting exponentials to these decays, the Rh+PATP decays (385 s)/(149 s) = 2.6 times faster than the reference PATP for visible illumination, while it decays (270 s)/(27 s) = 10 times faster for UV illumination, a factor of 10/2.6 = ~3.8 enhancement – again in excellent agreement with the 3.2 relative enhancement predicted by theory.

Figure 4 displays time-evolving Raman and fluorescence spectra for the Rh+PATP under constant 325 nm illumination at different sample locations than those used to obtain the spectra in Figure 3. Note that a fluorescence signal is also observed for UV illumination in the form of a growing baseline with a peak centered at 360 nm, located ~3000 $cm^{-1}$ from the UV excitation.[38] For most locations investigated, this fluorescence, integrated over a band from 2800-3280 $cm^{-1}$, also decayed with exposure time. As expected, fluorescence decayed faster for the Rh + PATP samples than for the reference PATP samples, and it decayed faster for UV illumination than for visible illumination. This indicates that the intense local field produced by excitation resonant with the LSPR is responsible for the accelerated decays of both the Raman and fluorescence spectra at most locations on the sample.

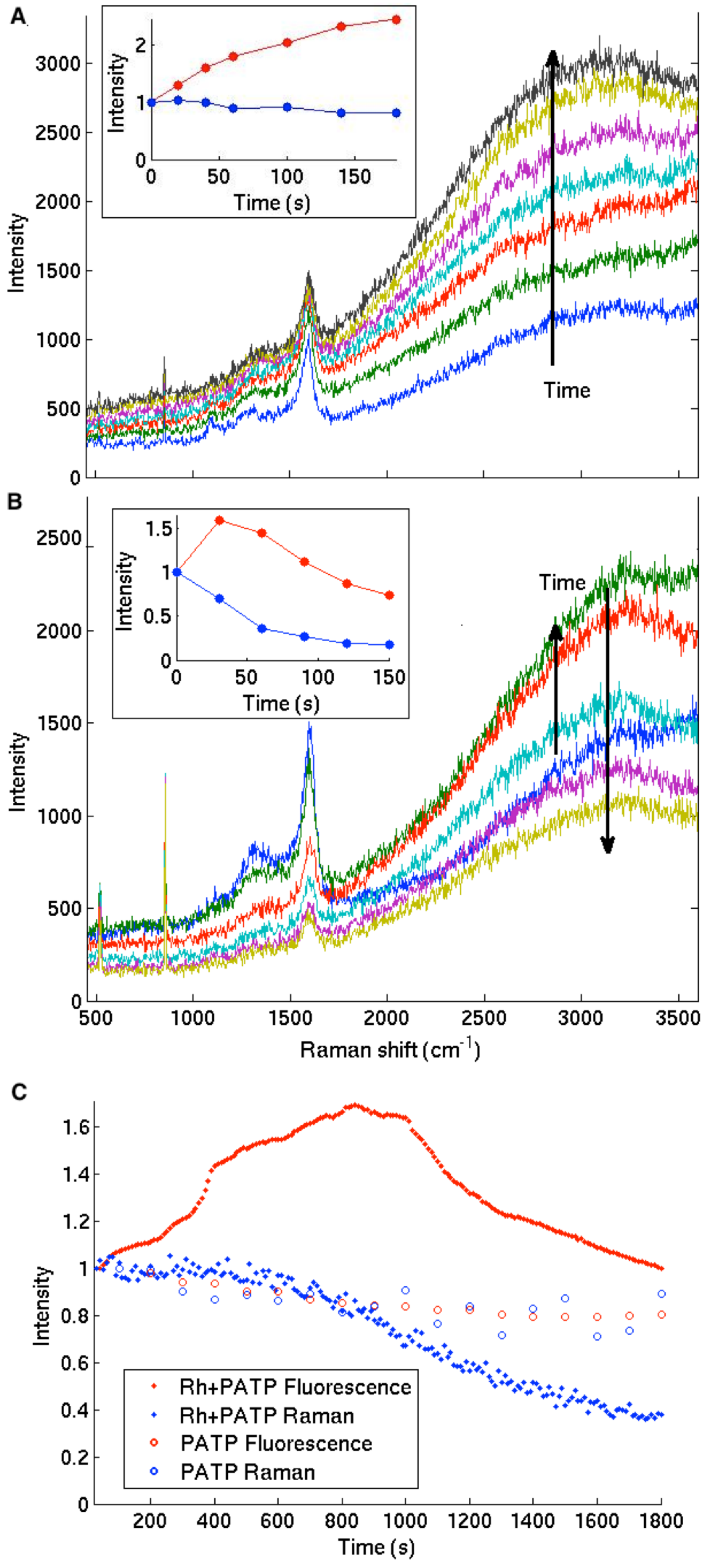

**Figure 4:** (a, b) Representative UV spectra of Rh+PATP, comparing the temporal evolution of the Raman and fluorescence spectra at two different illumination locations. The insets plot the temporal evolution of the 1600 $cm^{-1}$ Raman (blue) and 3000 $cm^{-1}$ fluorescence (red) features. (c) The strength of the fluorescence and Raman spectra from a third location as a function of time, compared to pure PATP, initialized to 1 at time t = 0.

However, charge transfer was also observed from the Rh NPs to PATP. Although unfavorable PATP alignment prevented us from observing the classic Raman fingerprint of enhanced $b_2$ vibrational modes previously shown to indicate charge transfer from Au NPs,[32,33] fluorescence brightening indicates that charge transfer is taking place. Site-to-site variations in the temporal behavior of the spectra reveal that the Raman feature in Figure 4 decays, but at different rates than in Figure 3, while the fluorescence feature grows, in one case for more than 800 s. This behavior is only seen in the Rh+PATP sample using a more tightly focused laser spot (110 $\mu m^2$ for Figure 4 vs 700 $\mu m^2$ for Figure 3) that allowed us to probe these site-to-site variations. As expected, charge transfer is more effective for these higher illumination intensities, and the site-dependent overlap of the illuminating field profile with the differing distributions of Rh+PATP in the interrogated regions likely produced the spatial variations observed.

Fitting exponentials to the decays in Figure 4 quantitatively reveals the varying behavior at different spots on the sample. In Figure 4a, the fluorescence rises sharply while the Raman feature decays very slowly with a time constant of 830 sec. Conversely, the Raman spectrum in Figure 4b decays rapidly with a time constant of 71 sec, and after a brief initial rise, the fluorescence also decays with a comparatively rapid time constant of 167 sec. These two characteristic behaviors represent opposite extremes of a range of observed plasmonically influenced decay: slow Raman decay indicating relatively stable PATP and rapid Raman decay indicating accelerated PATP photo-damage. Figure 4c represents an intermediate case in which PATP is initially stable, behaving as in Figure 4a where the Raman feature remains nearly constant as the fluorescence increases. After more than 13 minutes of illumination both Raman and fluorescence features begin to decay, but more slowly than in Figure 4b, with nearly identical time constants of 1300 and 1400 sec, respectively.

Clearly, different mechanisms are responsible for the NP-accelerated Raman spectral decay and the fluorescence brightening when the Raman signal is stable. We assert that only charge transfer from Rh NPs can explain PATP fluorescence brightening over such long timescales. Figure 4 indicates that when photo-damage of PATP begins, the Raman and fluorescence signals of the bare PATP decay similarly. Since C-C bonds within aromatic rings are most easily broken under intense UV illumination, the accelerated decay of the C-C Raman signal reflects local field-enhanced photo-damage of these rings near the Rh NPs. Until the PATP ring succumbs to photo-destruction following exposure to locally-enhanced UV irradiation, charge transfer may continue toward the amine group from the thiol linker that remains strongly attached to the Rh NP.

In conclusion, the UV plasmonic properties of Rh NPs were introduced by means of surface-enhanced Raman spectroscopy, surface-enhanced and charge transfer-enhanced fluorescence, and photo-induced degradation of the Raman dye PATP. Surface-enhanced spectroscopy was performed on a monolayer of PATP attached to chemically-synthesized Rh tripod nanostructures. By comparing the response for laser excitation resonant with (UV) and not resonant with (visible) the Rh NP LSPR, we found that Raman and fluorescence spectra were enhanced and photo-degradation was accelerated in the presence of Rh under resonant UV excitation. However, for more tightly focused UV illumination, fluorescence spectra often increased for many minutes, an indication that photoexcited hot electrons efficiently transferred from the Rh NPs to the attached PATP before photodamage ultimately quenched the fluorescence. These efficient photo-degradation and photo-induced charge transfer processes confirm the exciting potential of Rh nanostructures for UV plasmonic and photocatalytic applications.

Acknowledgements

The authors would like to thank Adam T. Roberts and Jay G. Simmons for facilitating the measurements and for useful discussions, and Dolores Ortiz for early contributions identifying promising UV plasmonic metals. This work has been supported by NSF-ECCS-12-32239. This work was partially supported by the Army's In-house Laboratory Innovative Research program. Financial support from USAITCA (project #W911NF-13-1-0245) and MICINN (Spanish Ministry of Science and Innovation, project #FIS2013-45854-P) is also acknowledged. AMW acknowledges support from the National Science Foundation Graduate Research Fellowship Program (Grant No. DGF1106401), and XZ acknowledges support from the Paul M. Gross fellowship from Department of Chemistry, Duke University. The authors declare no competing financial interests.

Author Information

Corresponding Author

* E-mail: everitt@phy.duke.edu

## Methods

**Rh NP synthesis.** The Rh nanoparticle tripods were synthesized by a modified polyol reduction method.[24] 4 mL of ethylene glycol (EG, J. T. Baker), held in a 50 mL round-bottom flask equipped with a reflux condenser, was heated to 140 °C under stirring and $N_2$. Into separate vials, each containing 1.6 mL EG, were dissolved $RhCl_3 \cdot xH_2O$ (38% Rh, 48.7 mg, 0.18 mmol) from Acros or poly(vinyl pyrrolidone) (PVP, Mr=55000, 102.6 mg, 0.9 mmol in terms of the repeating unit) from Aldrich. The $RhCl_3 \cdot xH_2O$ and PVP solutions were injected into the heated round-bottom flask simultaneously at a rate of 0.4 mL $min^{-1}$. After injection, the reaction mixture was kept at 140 °C for another 10 min to ensure complete reaction and then cooled to room temperature. The product was collected by centrifugation and washed by water and acetone several times. Finally, the Rh NPs were dispersed in ethanol to prepare the PVP-coated Rh suspension. The black precipitate was stirred with 10 mL of 1 M nitric acid at room temperature for 1 day to partially remove the PVP coating. The stripped Rh tripods were washed with water and acetone several times and dispersed in ethanol.

**Numerical methods.** Absorption calculations of figure 1b have been carried out using version 7.3.0 of the publicly available DDSCAT code.[39] DDSCAT uses the Discrete Dipole Approximation (DDA), also known as coupled dipole method.[26] Further details about assumptions and methodology can be found in Ref.[20]. The dielectric function for rhodium is plotted in Figure 5,[40] and the dielectric function for ethanol is from Ref. [41]. In order to simulate the UV-vis spectra of the nanoparticles in a liquid suspension, a total of 2000 snapshots representing random 3D rotations were used to evaluate the average cross-sections of the system:

a tripod star geometry composed of approximately 10,000 dipoles. Polydispersity was not considered since preliminary simulations under the assumption of single scattering were in good agreement with the experimental data. Finite-element simulations using the COMSOL MULTIPHYSICS code package[27] were used to map the field enhancements at 3.82eV (325nm) and 2.54eV (488nm) (Figure 1c). The inset of Figure 5 shows the tripod geometry, a threefold symmetric star, on a 1 μm-thick $SiO_2$ “glass” substrate with refractive index n = 1.5.[40] In order to calculate the UV/visible enhancement ratio, the system was illuminated under normal incidence with a fixed linear polarization of the lasers. Then, $|E|^2$ on the surface was averaged over all orientations of the Rh tripod around the substrate normal axis.

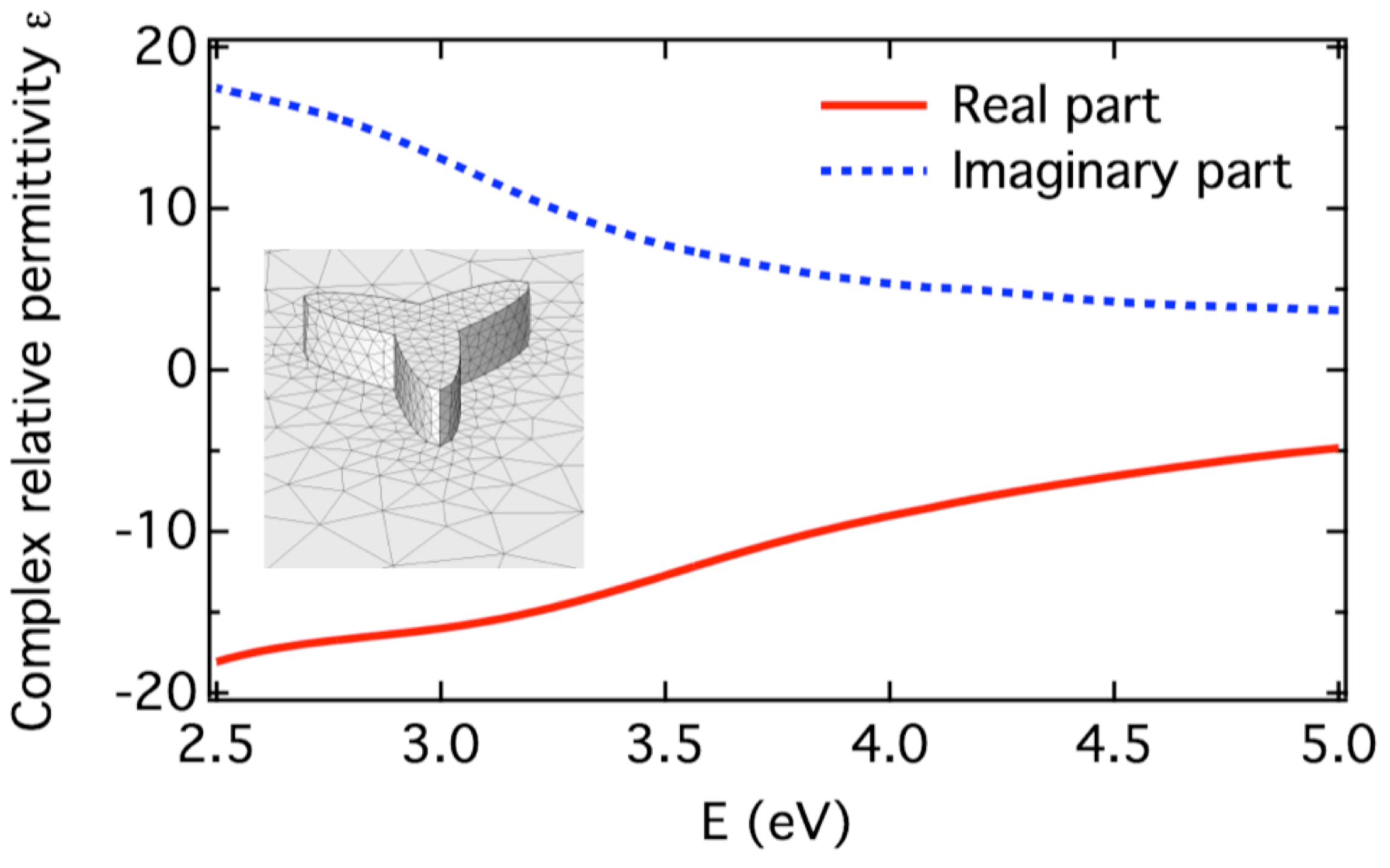

**Figure 5:** Dielectric function for Rh used in the calculations. The inset shows the mesh used for COMSOL simulations of the Rh tripod on an $SiO_2$ “glass” substrate.